# Three months journeying of a Hawaiian monk seal

**David R. Brillinger**[*1], **Brent S. Stewart**[†2] **and Charles L. Littnan**[3]

*University of California, Berkeley, Hubbs-SeaWorld Research Institute and Pacific Islands Fisheries Science Center*

**Abstract:** Hawaiian monk seals (*Monachus schauinslandi*) are endemic to the Hawaiian Islands and are the most endangered species of marine mammal that lives entirely within the jurisdiction of the United States. The species numbers around 1300 and has been declining owing, among other things, to poor juvenile survival which is evidently related to poor foraging success. Consequently, data have been collected recently on the foraging habitats, movements, and behaviors of monk seals throughout the Northwestern and main Hawaiian Islands.

Our work here is directed to exploring a data set located in a relatively shallow offshore submerged bank (Penguin Bank) in our search of a model for a seal's journey. The work ends by fitting a stochastic differential equation (SDE) that mimics some aspects of the behavior of seals by working with location data collected for one seal. The SDE is found by developing a time varying potential function with two points of attraction. The times of location are irregularly spaced and not close together geographically, leading to some difficulties of interpretation. Synthetic plots generated using the model are employed to assess its reasonableness spatially and temporally. One aspect is that the animal stays mainly southwest of Molokai. The work led to the estimation of the lengths and locations of the seal's foraging trips.

## 1. Introduction

This paper studies a three month journey of a juvenile female (4-5 years old) Hawaiian monk seal, (*Monachus schauinslandi*) while she foraged and occasionally hauled out ashore. She was tagged and released at the southwest corner of Molokai, see Figure 1. (The animal was tracked from 13 April 2004 through 27 July 2004.) She had a satellite-linked radio transmitter glued to her dorsal pelage to document geographic and vertical movements as proxies of foraging behavior.

An important reason for studying the Hawaiian monk seal is that this is the most endangered species of marine mammal living entirely within United States

[*]Supported by NSF Grants DMS-05-04162 and DMS-07-07157.

[†]Supported by contracts from NOAA, PIFSC and grants from Hubbs-SeaWorld Institute.

[1]University of California, Department of Statistics, 367 Evans Hall #3860, Berkeley, CA 94720-3860, USA, e-mail: brill@stat.berkeley.edu

[2]Hubbs-SeaWorld Research Institute, 2595 Ingraham Street, San Diego, CA 92109, USA, e-mail: stewartb@hswri.org

[3]Pacific Islands Fisheries Science Center, NOAA Fisheries, 2570 Dole Street, Honolulu, HI 96822, USA, e-mail: charles.littnan@noaa.gov

246

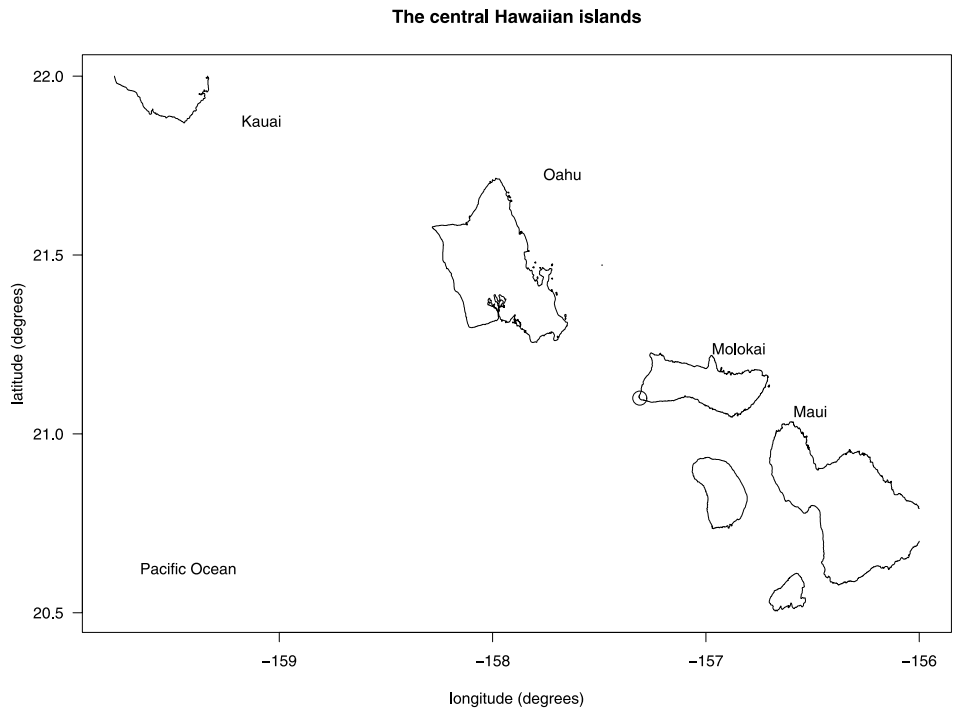

FIG 1. *The central Hawaiian Islands. The circle on the southwest corner of Molokai is La'au Point, the location where the seal was initially captured.*

jurisdiction. Until relatively recently, little was known about the biology and ecology of this species as their remote location and limited technology for studying them prevented informative study [33].

The species now numbers around 1300 seals, principally in the Northwestern Hawaiian Islands but with increasing numbers in the main Hawaiian Islands [2, 20, 35]. The key factor in the recent decline of Hawaiian monk seals is poor survival of juvenile seals, hypothesized to be related to poor foraging success. Consequently, the basic motivation of contemporary data collection is to learn where the animals go to forage geographically and vertically. Such information is needed for management and conservation purposes. This work studies the geographic movements of a single animal preparatory to analysis of a larger data set.

As the data stand, one can answer the management question as to where the seals go, but the techniques being developed here refine understanding of the data, help better understand how the animals move in their environment, and increase their value to management. Statistical analysis can help in addressing such questions, provided pertinent data are available. In particular, methods are needed to describe associations between movements and various potential physical and biological explanatory variables.

Statistical aspects include:

How to display, process, and describe such data?

How to include explanatory variables?

How to highlight unusual aspects?

How to obtain representative tracks for an animal?

How to interpolate between available time points?

The analyses presented combine both space and time. Related statistical work on the tracks of northern elephant seals, and of elk and deer was developed and described in [9]. The models employed there included stochastic differential equations (SDEs) for drift on a sphere and for motion on the plane with a spatially dependent drift term, as well as including other explanatory variables beyond space and time. Also to be mentioned are the papers [11, 17] which handle difficulties resulting from measurement errors. The second paper is in part motivated by [18] as was this work.

The initial analyses employ robust-resistant tools to explore the situation, for there is substantial measurement error present. The later analyses employ a "cleaned" data set and develop a stochastic model.

The objective is really post-hoc, directed at determining if exploratory statistical techniques and stochastic differential equation models might help address general questions using this particular novel data set. The statistical goal is obtaining a model that mimics the behavior observed in the data set.

## 2. Mise en scène

The focus of the work is both exploratory, looking for insight and understanding in a novel data set, and formal, developing a model for an entity living and moving in space, and time.

Studying motion, and the statistics of motion, has a long and venerable history. To begin, one can note the Newtonian equations of motion:

$$d\mathbf{r}(t) = \mathbf{v}(t)dt,$$

$$d\mathbf{v}(t) = -\beta\mathbf{v}(t)dt - \beta\nabla H(\mathbf{r}(t), t)dt$$

with $\mathbf{r}(t)$ a particle's location at time $t$, $\mathbf{v}(t)$ its velocity, $H(\mathbf{r}, t)$ a potential function, and $\nabla$ the gradient. The potential function controls a particle's direction and velocity. Regions of attraction and repulsion, for example, may be introduced by terms in $H$. The parameter $\beta$ represents friction. One reference is [23].

In the case that $\beta$ is large, the equations become, approximately,

$$d\mathbf{r}(t) = -\nabla H(\mathbf{r}(t), t)dt.$$

If one adds a stochastic term, and changes the notation slightly, then one obtains the stochastic differential equation

(1) $$d\mathbf{r}(t) = \boldsymbol{\mu}(\mathbf{r}(t), t)dt + \boldsymbol{\Sigma}(\mathbf{r}(t), t)d\mathbf{B}(t)$$

having written $\boldsymbol{\mu} = -\nabla H$. Often $\mathbf{B}(t)$ is assumed to be Brownian motion. In the planar case with $\mathbf{r} = (x, y)$, the 2-vector $\boldsymbol{\mu}$ contains the partial derivatives $H_x, H_y$. A unique solution exists under regularity conditions on $\boldsymbol{\mu}$ and $\boldsymbol{\Sigma}$. In the stationary case, with $\boldsymbol{\Sigma} = \sigma_0 \mathbf{I}$, there may be an invariant density given by

$$\pi(\mathbf{r}) = c \exp\{-2H(\mathbf{r})/\sigma_0^2\}, \qquad H(\mathbf{r}) = -\frac{1}{2}\sigma_0^2 \log \pi(\mathbf{r})/c$$

with $c$ an integration constant. There is an important advantage to working with $H$ in the modelling, namely it is scalar-valued, whereas $\boldsymbol{\mu}$ is vector-valued.

A long standing SDE, motivated by motion, is that of Ornstein and Uhlenbeck. It has the parameterization

$$\boldsymbol{\mu}(\mathbf{r}, t) = \mathbf{A}(\mathbf{a} - \mathbf{r}) \qquad \text{and} \qquad \boldsymbol{\Sigma}(\mathbf{r}(t), t) = \boldsymbol{\Sigma}$$

with **a** a point of attraction to which the moving object is lead. The potential function here is

$$H(\mathbf{r}) \;=\; (\mathbf{a}-\mathbf{r})^{\tau}\mathbf{A}(\mathbf{a}-\mathbf{r})/2,$$

page 170 in [1].

Another pertinent planar SDE, due to Kendall [18], is

$$
\begin{aligned}
dr(t) &= \left(\frac{\sigma^2}{2r(t)} \,-\, \delta\right)dt \,+\, \sigma dU(t), \\
d\phi(t) &= \frac{\sigma}{r(t)}dV(t)
\end{aligned}
\tag{2}
$$

with $U$, $V$ independent standard Brownians. This was proposed to model bird motion towards an origin. Here $r(t)$ is the bird's distance from the origin at time $t$ and $\phi(t)$ the bird's direction. The $\delta$ here represents speed towards the origin. The potential function for this case is

$$H(r,\phi) \;=\; \frac{\sigma^2}{2}\log\; r \,-\, \delta r. \tag{3}$$

Writing this function in Cartesian coordinates, $\mathbf{r} = (r_1, r_2)$, and supposing that **a** is a point of attraction, if one computes the gradient with $\boldsymbol{\mu} = (\mu_1, \mu_2)$ then one obtains the drift components

$$\mu_j \;=\; \delta(r_j - a_j)/\|\mathbf{r}-\mathbf{a}\| \,-\, \frac{\sigma^2}{2}(r_j - a_j)/\|\mathbf{r}-\mathbf{a}\|^2 \;, \quad j = 1, 2. \tag{4}$$

Using the multi-dimensional Ito formula (see [26], Section 4.2) with the destination **a**, Kendall's SDE becomes, in Cartesian coordinates,

$$d\mathbf{r}(t) \;=\; \boldsymbol{\mu}(\mathbf{r}(t))dt \,+\, \sigma d\mathbf{B}(t) \tag{5}$$

with $\boldsymbol{\mu}$ given by (4) and with **B** bivariate Brownian motion. This is the model that will be employed in the data analyses, but $H$ will depend on $t$ as there will be two time periods each with their own point of attraction.

The computations using $H$ show an advantage of the potential function approach, namely for the drift term $\boldsymbol{\mu}$ the switch from polar to Cartesian coordinates is direct, whereas the use of Ito's formula requires some careful algebra.

Kendall's model was generalized in [3] to motion on the sphere and to the equations

$$
\begin{aligned}
d\theta(t) &= \left(\frac{\sigma^2}{2\tan\;\theta(t)} \,-\, \delta\right)dt \,+\, \sigma dU(t), \\
d\phi(t) &= \frac{\sigma}{\sin\;\theta(t)}dV(t).
\end{aligned}
$$

This model was employed to describe the lengthy migrations of Northern elephant seals in the Pacific Ocean. Here $\theta(t)$ and $\phi(t)$ correspond to colatitude and longitude respectively.

Further discussion and examples of the potential approach may be found in [5, 6, 8]. The case of motion in a region with a boundary is reviewed in [4].

## 3. Some ecological questions to address

Understanding the foraging behavior and habitat use of Hawaiian monk seals is critical to understand the continued decline of the species. There are several specific questions that are relevant to this end e.g. [10, 21, 27, 28, 34]:

What are the geographic and vertical marine habitats that the Hawaiian monk seals use?

What habitats are essential, with some buffer, to the survival and vitality of this species?

Are there age and sex differences in the habitats they use when foraging?

Do seals have individual preferences in foraging locations and does an individual vary its behavior over different time scales?

How long is a foraging trip?

## 4. The data

The experiment begins with a satellite-linked time depth recorder (SLTDR) being glued onto the seal's dorsal pelage. The SLTDR records times and depths of dives and transmits a brief radio frequency signal to a system of near polar earth orbiting satellites managed by the Argos Data Collection and Location Service. Periodic locations of the seal being studied were determined by measurements of Doppler shifts in the reception of successive transmissions. Those locations and accompanying dive data were then communicated daily to the marine biologists by email.

Attached to a location estimate is a prediction of the location's error (LC or location class). The index takes on the values 3, 2, 1, 0, A, B, Z. When LC = 3, 2, or 1 the error in the location is predicted to be 1 km or less. Errors for locations of LC 0, A, and B are not predicted by Argos but may often be around 1 km but up to 10 to 20 km on occasion. In this paper when LC is 3, 2, 1, 0, a location is referred to as well-determined.

The principal data to be employed in the analyses to be presented are the locations estimated at various known times and their LC values. The locations are determined several times a day. In total there were 573 locations (see Figure 2) and the number of well-determined, i.e. LC better than A, was 189 (see Figure 3).

The location values employed in the computations are in UTM (universal transverse mercator) coordinates. The units are metric (meters and kilometers). To describe the UTM system think of wrapping the earth, not by a cylinder touching the earth's equator as in the Mercator system, rather with a cylinder touching at a more convenient place given the data set's locations. The earth is then projected onto that cylinder and UTM coordinates determined. Using UTM coordinates has the advantage of preserving distances, providing a better projection into a plane and avoiding the use of spherical statistics. If distances are not small, use of latitude-longitude coordinates can lead to biases in analyses. The use of UTM coordinates makes it easier to digest information like: the animal's usual speed is between 1.5 and 2.5 m/sec. One can convert back and forth between latitude-longitude and UTM coordinates using basic programs.

The data employed covered 85 days, with the time of location varying irregularly and no estimates for some time periods because of the satellite's orbit. Figure 2 shows all 573 locations of the data set. One sees clusters of locations to the southwest (SW) of Molokai. Also there are quite a number of broadly scattered locations.

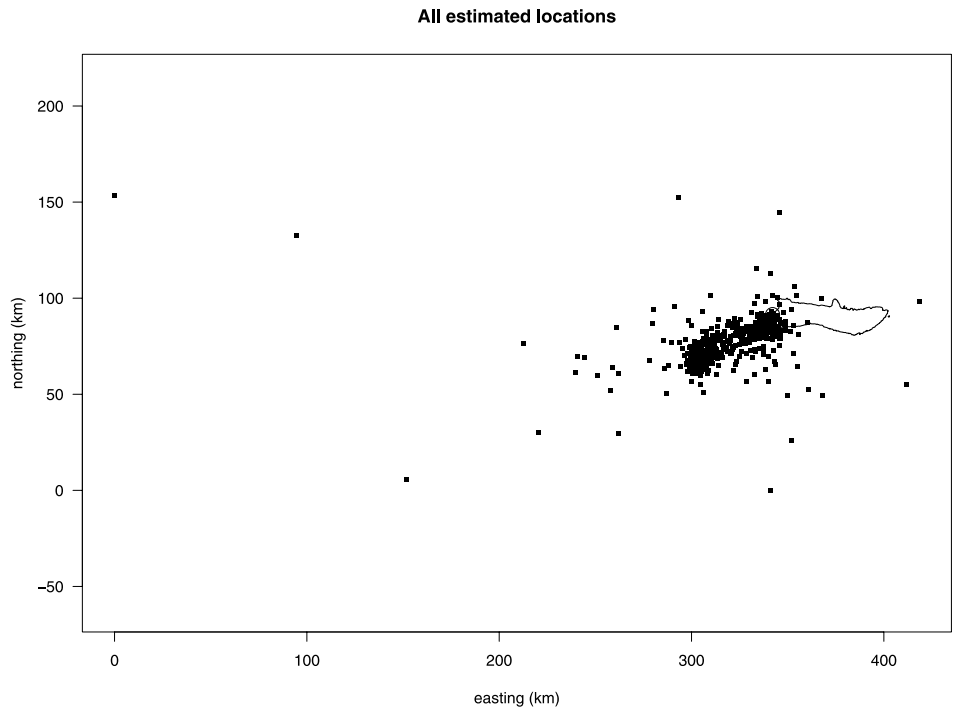

FIG 2. *The figure shows all 573 of the estimated locations. The island is Molokai. The circle indicates the initial estimated position. The coordinates are UTM.*

Figure 3 shows the 189 well-determined locations, i.e. those with LC = 3, 2, 1, 0. One sees a mass of locations to the SW of Molokai. The 200 fathom line has been added to the figure. The well-determined locations of the seal are nearly all within this boundary, with a few on the land! This latter occurrence is reflective of the measurement error involved. The 200 fathom region is roughly Penguin Bank, a relatively broad shallow bank extending south west from Molokai. The region is part of a marine sanctuary for humpback whales.

A concern is how to describe the general journeying of the seal. One can expect the animal to be continually searching out food and safe habitat to rest. Figure 4 is an attempt to learn something of the basic flow of the seal's movement. The sampled trajectory is split into 4 segments. Initially the animal is seen to remain close to Molokai. The subsequent periods show the animal ranging broadly within Penguin Bank. The track shows the animal meandering at some distance from the initial point.

One notes that the animal apparently takes some lengthy excursions out of the 200 fathom region.

## 5. Some statistical background

To begin, a novel robust-resistant statistical tool, the *bagplot* [30] will be employed. The bagplot can be helpful to highlight unusual data points and structure in multidimensional data. It proves useful here because of the presence of substantial location errors.

**The well–determined points**

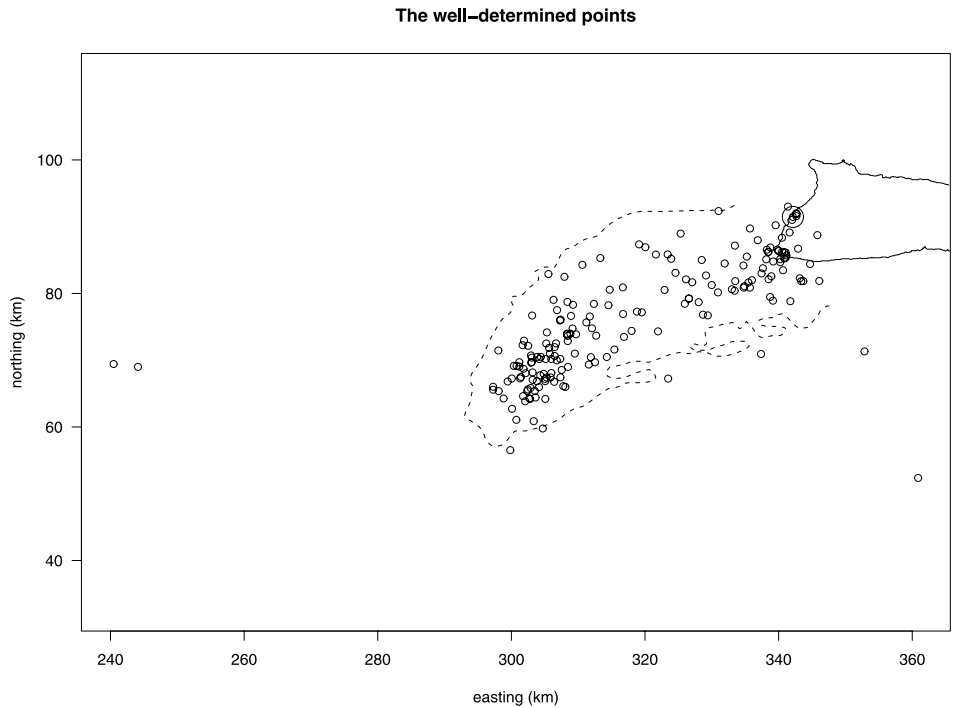

Fig 3. *The figure shows the 189 well-determined locations. The dashed line is the 200 fathom line representing Penguin Bank. The large circle is the initial estimated location. There are some locations well inland suggesting the measurement error is greater than the claimed 1 km.*

The bagplot is a multi-dimensional generalization of the boxplot. Its center is a multidimensional median. Its "bag" contains 50% of the observations with greatest depth and generalizes the box of the boxplot. The bagplot's "fence" separates inliers from outliers as does the boxplot's fence. Observations between the bag and the fence are marked by whiskers. The fence is obtained by inflating the bag, from the center, by a factor of 3.

The bagplot is equivariant under affine transformations unlike various other multivariate displays that have been proposed. It provides information on location, spread, association, skewness, tails and outliers of multidimensional data. For an example see Figure 5. The R program employed comes from [37]. In Figure 6, a sliding bagplot helps understand the seal's spatial-temporal motion.

Stochastic differential equations such as (1) have already been referred to. There has been a substantial amount of work on statistical inference for SDEs, references include [15, 32]. One can motivate later inferential work by setting down the so-called Euler approximation

$$(6) \qquad \mathbf{r}(t_{i+1}) - \mathbf{r}(t_i) \approx \boldsymbol{\mu}(\mathbf{r}(t_i), t_i)(t_{i+1} - t_i) + \boldsymbol{\Sigma}(\mathbf{r}(t_i), t_i)\mathbf{Z}_i\sqrt{t_{i+1} - t_i}$$

with the $t_i$ an increasing sequence of time points filling in the time domain of the problem [see 19], and the $\mathbf{Z}_i$ independent bivariate standard normals. The $t_i$ may be thought of as the times of observation in the present work. Approximations such as this may be used to show that solutions exist for various SDEs [19].

If one is willing to assume that $\boldsymbol{\mu}(\mathbf{r}, t)$ is a smooth function of $(\mathbf{r}, t)$, one can use nonparametric estimation methods [8] to estimate it. Assuming that $\boldsymbol{\mu}(\mathbf{r}, t) = \boldsymbol{\mu}(\mathbf{r})$

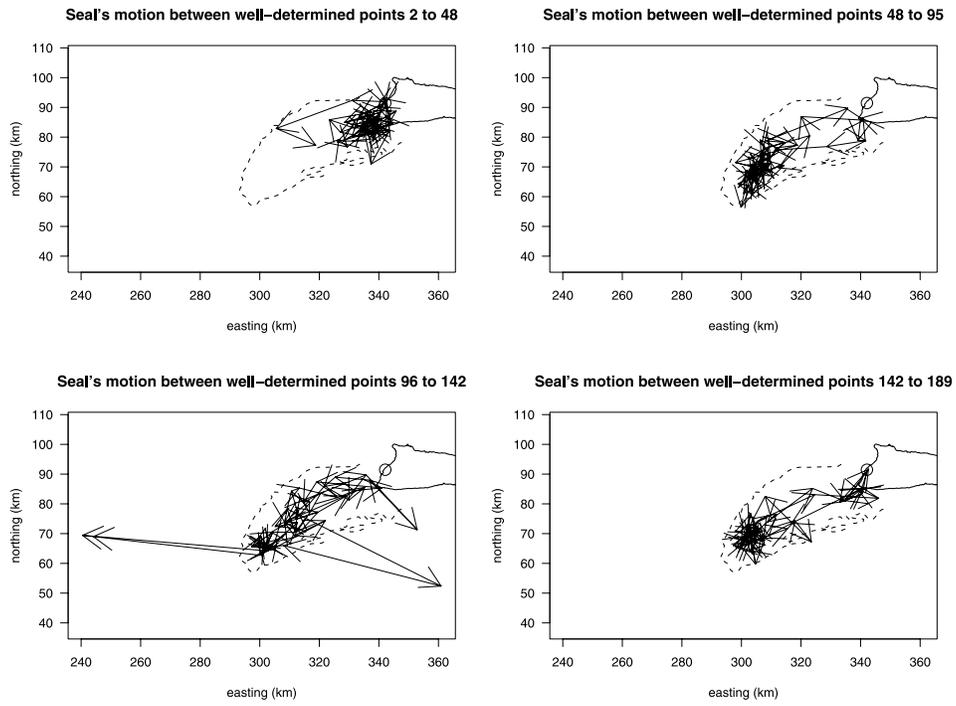

FIG 4. *Directed arrows representing the motion in four contiguous segments. The data employed are the well-determined locations. The track starts at the large circle.*

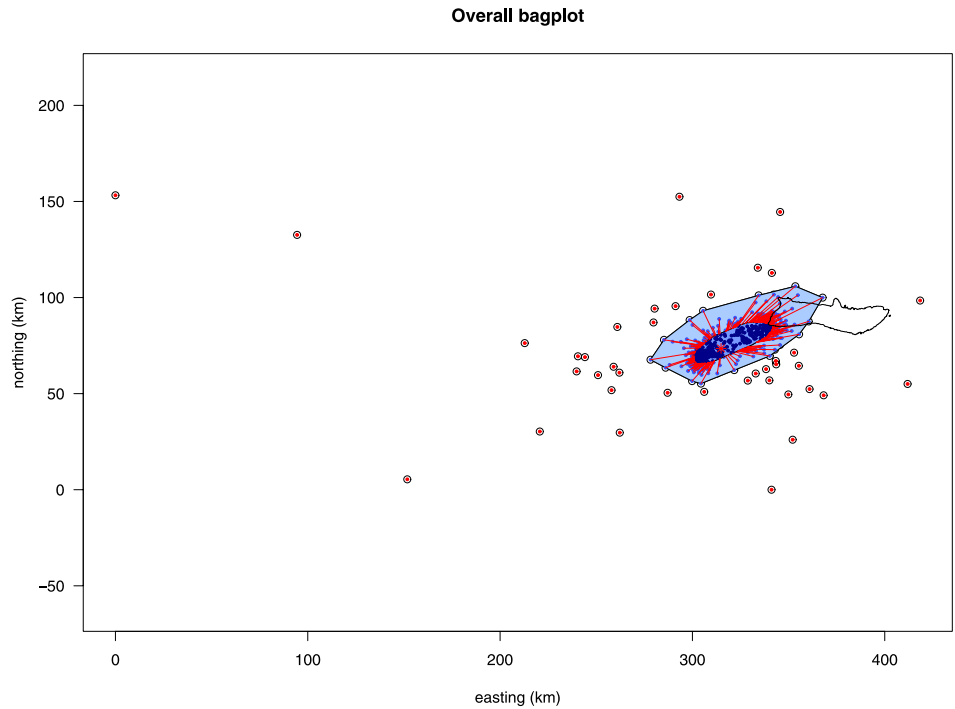

FIG 5. *Bagplot of all the estimated locations. The bag and fence add detail to Figure 2.*

Bagplots for six successive time periods and $\boldsymbol{\Sigma}(\mathbf{r}(t), t) = \sigma \mathbf{I}$, one can consider $\hat{\sigma}^2$ as an estimate of $\sigma$, where

$$(7) \qquad 2\hat{\sigma}^2 = \frac{1}{I} \sum_i ||\mathbf{r}(t_{i+1} - \mathbf{r}(t_i) - \hat{\boldsymbol{\mu}}(\mathbf{r}(t_i))(t_{i+1} - t_i)||^2 / (t_{i+1} - t_i),$$

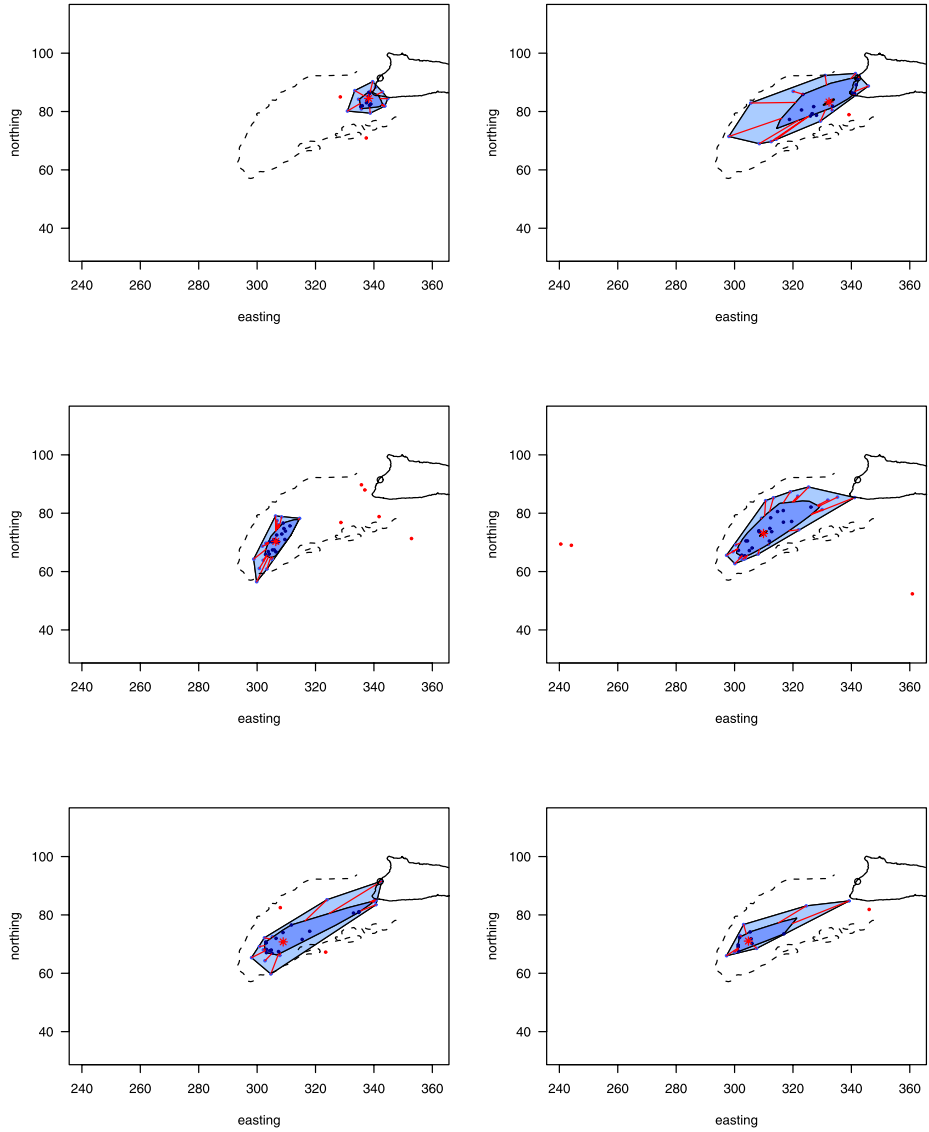

FIG 6. *Bagplots of the well-determined locations for six successive 14-day periods. One reads the panels from left to right, dropping down the rows. The 200 fathom line has been added in each case.*

$i = 1, \ldots, I$. In the present work, a further aspect is that the motion is practically bounded in space.

In [4] various methods of simulating an SDE moving in a bounded region are reviewed. What will be done in the work here is that (6) will be employed to compute a possible $\mathbf{r}(t_{i+1})$ given $\mathbf{r}(t_i)$, $\boldsymbol{\mu}$ and independent normal variates. If that point falls outside the region then it will be pulled back to the closest point on the boundary. This method seems a reasonable approximation to a solution for small $dt$, see [22, 29].

Creating a stochastic model can lead to empirical insight and understanding. In particular one can move onto other things, synthetic plots and bootstrapping for example. These involve the simulation of realizations of the model. Comparing synthetic plots to the data plots at hand allows assessment of reasonableness. Synthetic plots were employed in Neyman et al [25]. The paper [24] speaks of "a novel method of testing certain cosmological models." After generating their synthetic plots these researchers were led to change a proposed model for the distribution of galaxies.

In the "Turing test" [31], the data and the simulated data are given to a referee who are invited to tell which is the actual data.

Here the modified version of expression (6) will be used in the simulations. The modification is needed because of the region provided by Penguin Bank. This region can be approximated by the 200 fathom line and appears as the dashed line in Figure 3.

## 6. Further descriptive analyses

Figure 5 provides the bagplot of all 573 of the data points, the well-measured and the not. It may be compared directly with Figure 2, the difference being that a bagplot has been added. The bagplot shows where the measurements are centered in a robust sense, and suggests substantial outliers and also, via the whiskers, points to be reviewed. The central convex region, contains 50% of the data points.

The elongation of the bag and fence regions in Figure 5 extends to the southwest. These features can be discerned by looking at Figure 2 alone, but are apparent by a glance at Figure 5 with the bagplot added.

Figure 6 is based on the well-determined values only. It provides bagplots for six contiguous 14-day time periods and provides more information about the motion and the presence of outliers. The top left panel suggests that the seal remained close to where it was introduced into the water after being instrumented. This is consistent with Figure 4. The point of release is indicated by the larger circle on the coast. The top right panel in Figure 6 shows the seal expanding its range principally to the southwest with a broadening in the NW-SE directions too. The middle left panel suggests that the seal moved its center of location offshore. The final three panels show the seal's motion centered offshore while ranging in a SW-NE direction. Due to the vagaries of the observation times, these plots are based on 25, 32, 39, 44, 32, 16 points respectively. The bags are skinny suggesting a preferred line of movement. The data will be further separated in time in Figure 7.

A reason for using UTM coordinates here is that it avoids the development of a spherical bagplot. The UTM coordinates provide a better approximation when using linear methods.

Now there is a final cleaning of the data set. Specifically, points such that the animal would have to swim at a speed of more than 3 m/sec (259.2 km/day) to get to the next location are eliminated. The resulting data set contains 173 points and will be called the cleaned data set. It will be used throughout the remaining analyses.

Figure 7 attempts to understand the temporal behavior further. Since the animal appears to return to the neighborhood of La'au Point from time to time, the figure graphs the distance of a seal location from La'au Point versus the time. An interesting structure appears in the figure. Initially, for up to 20 days after the start of the measurements, the animal remains close to La'au Point. Then she heads some distance out to sea followed by a return to near La'au Point. This continues for perhaps four foraging trips, namely the segments 4, 6, 7, 9 in the figure. These segments were chosen by eye.

Figure 8 shows the four foraging trips suggested by the vertical lines in Figure 7. The panels show common components of behavior. Their temporal lengths are respectively 15.70, 9.68, 12.18, 8.33 days. The outlier in the top right panel could not be ruled out by location class, LC, or speed considerations.

## 7. Fitting an SDE model

Figures 7 and 8 prove useful for developing a stochastic model. The model

$$d\mathbf{r}(t) \; = \; \boldsymbol{\mu}(\mathbf{r}(t))dt \; + \; \sigma d\mathbf{B}(t), \qquad \mathbf{r}(t) \; \in \; F$$

will be employed, with $\mathbf{B}$ Brownian, and with the motion restricted to within $F$, the 200 fathom region.

The exploratory analyses and particularly Figures 7 and 8 suggest considering a model with two points of attraction for the animal; a point at the southwest end of the 200 fathom region for the outbound journey and foraging and another at La'au Point for the inbound journey. The potential function restricted to the 200 fathom region will be employed to introduce these points. The function, $H$ of Section 2, will change depending on whether the animal is in the outbound-foraging or the inbound stage. Specifically, it will depend on both location, **r**, and time, $t$. There are 27 observations in all.

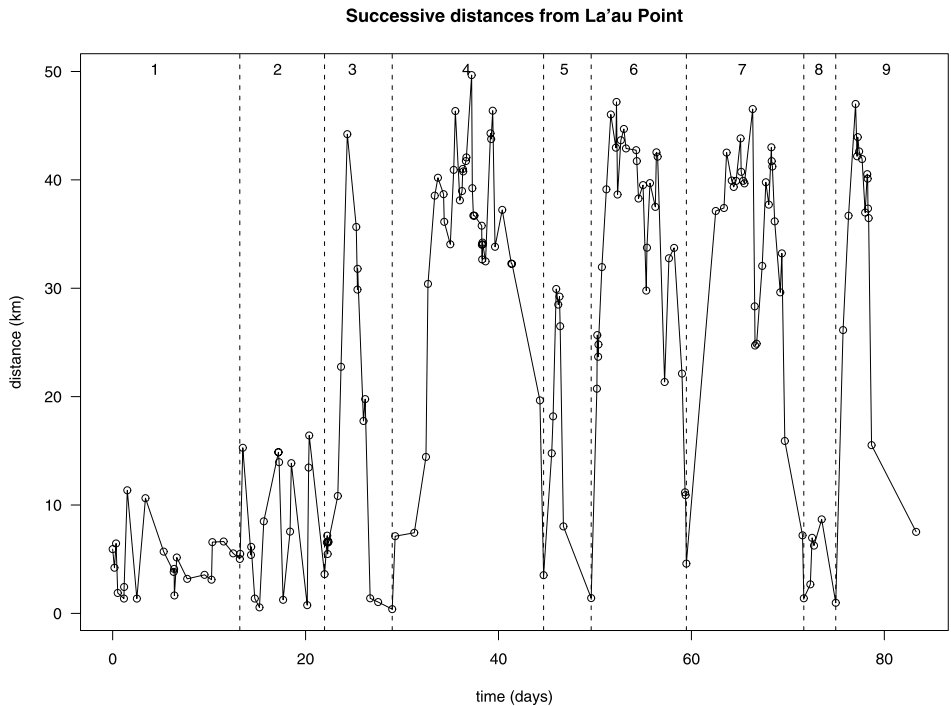

FIG 7. *The distance of the seal from La'au Point plotted against the time of location. The vertical lines delineate apparent journeys.*

The $\boldsymbol{\mu}$ function of (4) was fit to the cleaned data set. There are different points of attraction for the outbound and inbound journeys. The outbound point was determined by eye as (302.8, 67.5). See the SW circle in Figure 9. The inbound was La'au Point.

Robust linear regression, with weights $\sqrt{t_{i+1} - t_i}$, was used to determine estimates of the parameters. A robust procedure was used because of the sometimes large measurement errors in the location estimates. It is assumed that the destination and transition are known in the computation. The results are presented in Table 1.

The parameter $|\delta|$ represents speed towards the attractors. Here the estimate is 17.2 km per day, with an s.e. of 3.8.

Table 1 essentially contains two estimates of $\sigma$ (lines 3 and 4). The difference possibly represents measurement error on the positions. Measurement error is included,

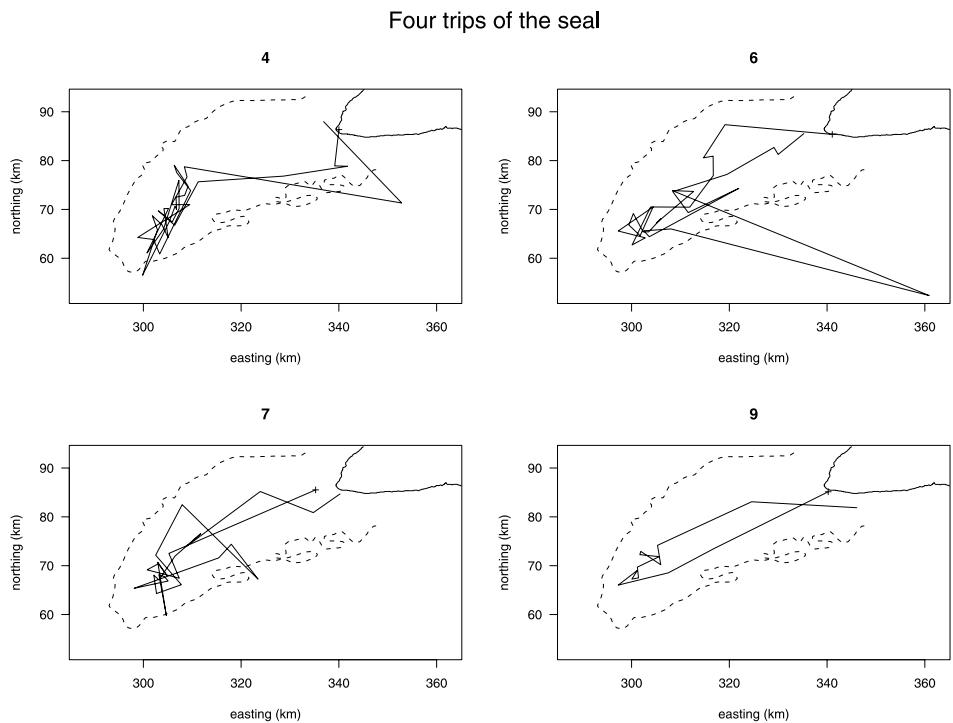

Four trips of the seal

FIG 8. *The separate plots display the journeys 4, 6, 7, and 9 delineated in Figure 7. The initial point of a trip is indicated by "+".*

TABLE 1

| Parameter | Estimate | s.e. |
|---|---|---|
| $\delta$ | −17.2 km/day | 3.8 km/day |
| $\sigma^2/2$ | 11.3 | 9.2 |
| $\sigma$ | 12.2 | |
| Destination | (302.8, 67.5) | Eye estimate |
| Transition | 9.8 days | |

**Potential function for outbound journey**

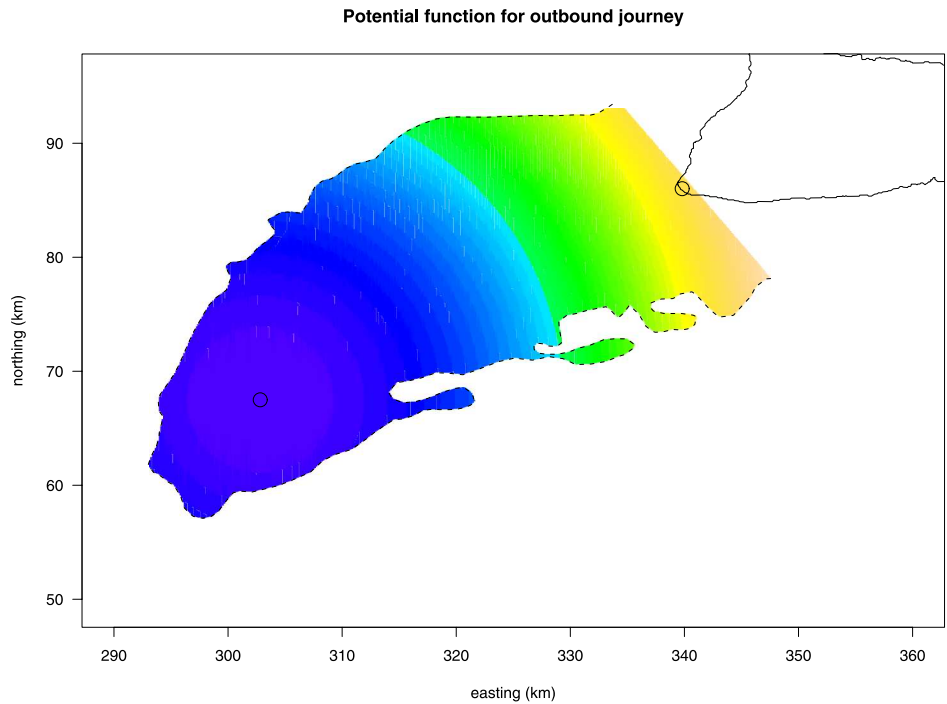

FIG 9. *The potential function employed to describe the outbound and foraging journeys. The circle in the SW corner represents an assumed point of destination. Lighter values correspond to large and darker to smaller potential function values. The seal is attracted to the darker region.*

and fit, in a related model for elephant-seal motion [7].

The indicated standard errors were output by the function rlm() of the MASS library of [16, 36].

## 8. Synthetic plots

Figure 9 shows the estimated potential function of (3) expressed in Cartesian coordinates. It has an attractor at (302.8, 67.5) and slopes downwards from La'au Point to the southwest. It leads to the particle being pulled outward to the southwest.

Figure 10 shows the estimated function for the return journey to La'au Point. It has La'au Point as the attractor and slopes in that direction.

Figure 11 presents a synthetic journey developed using the parameter values of Table 1 and $dt = 1$ hr. The time spacings were those of journey 7, hence several long stretches. The journey is taken to start at La'au Point. The path may be compared to those of Figure 8, particularly journey 7.

Figure 12 is the analog of Figure 7. It shows a particle starting near La'au Point, heading off about 50 km, foraging and then heading back to Lau'au Point with the journey taking about two and a half days.

Measurement error might have been included. Perhaps a permutation of the residuals might have been added.

**Potential function for inbound journey**

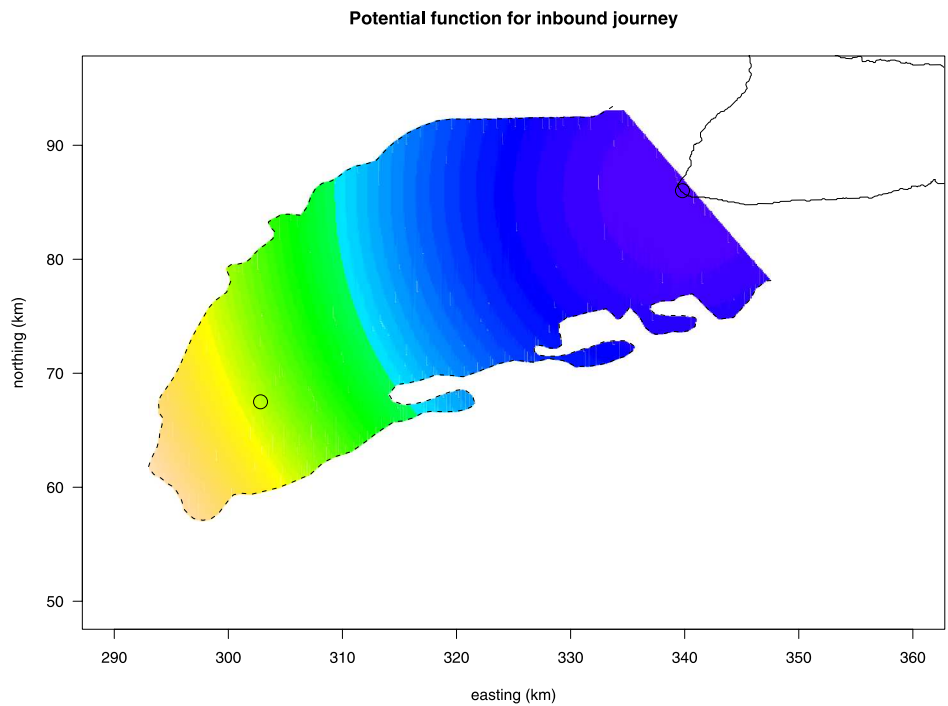

FIG 10. *The potential function employed to describe the inbound journey. The circle on Molokai represents La'au Point. Again the animal is pulled towards the darker region.*

## 9. Assessment

Stochastic models, once validated, have important uses: detecting change, prediction, sensitivity analysis,... The results themselves are useful for: comparisons, developing base values, and estimates of biological parameters (eg. speed) amongst other things.

On page 193 in Freedman [12] one finds the following stricture:

> Assumptions should be made explicit. It should be made clear which assumptions were checked and how the checking was done. It should also be made clear which assumptions were not checked.

1. The assumptions were explicit, specifically it was assumed that the particle followed a trajectory generated by the model (6) and the data values resulted from sampling it at the times of journey 7.
2. The assumptions were checked by constructing a synthetic plot.
3. The Brownian assumption was not checked.

## 10. Difficulties of interpretation

There are important difficulties in seeking to interpret the data and the results of the analyses presented. To start, the study is observational. For a discussion of the problems with observational studies see [12].

A second difficulty is that the animal's track is not observed continuously, rather estimated locations are available only at scattered times. This leads to unavoidable difficulty in interpreting the results, for the animal might go slowly directly between

**Synthetic plot of a journey**

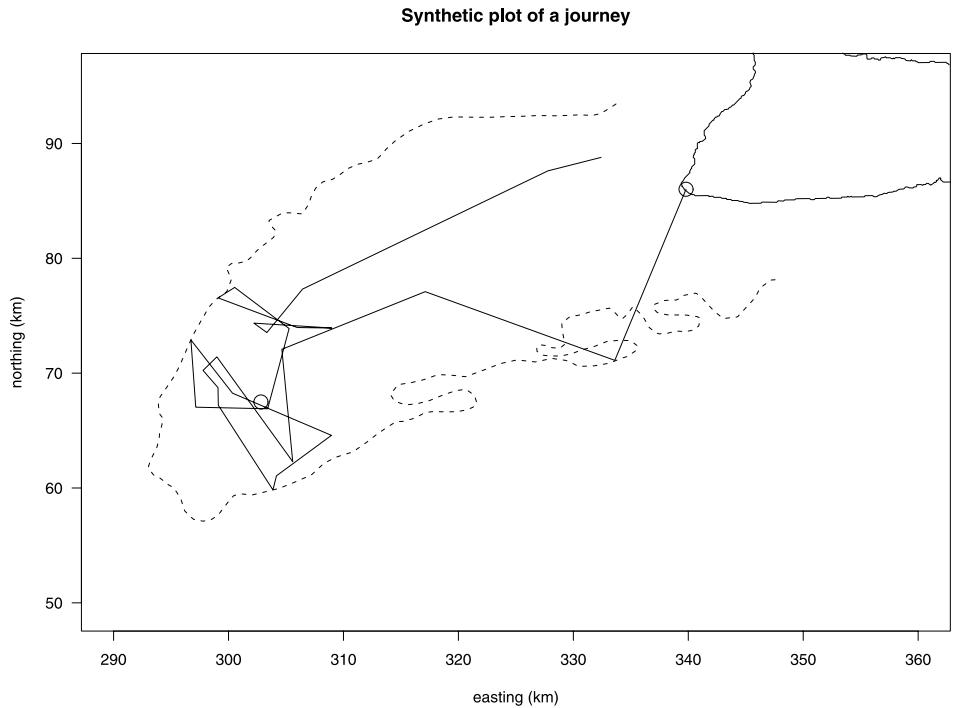

FIG 11. *A simulated journey starting at La'au Point and heading towards the circle at the lower left. The potential function of Figure 9 is employed. This figure may be compared with Figure 8. The kinks in the track are the time points of journey* 7 *of Figure* 7.

two locations or it might go very quickly in a round about way between the points. Further, in this study the time differences between successive observations are large and therefore it can be difficult to recognize the character of the motion. There is no simple way out of these problems. With animals on land one can make readings at finely spaced time points, but at sea, the animal may be diving and no transmissions possible for lengthy periods.

Continuing, some parameters are estimated by eye, a method fraught with difficulties. However, this is a preliminary study.

## 11. Discussion

This work constitutes an exploratory analysis, but there are data to be had from other animals to re-examine the results. The work searched out a stochastic model using descriptive methods, classical dynamics and statistical techniques. Developing a pertinent potential function proved an effective manner by which to infer a model. There was preliminary model assessment by looking at pictures of simulations and also some residual analysis.

The computations employed the statistical package R [16], including bagplot displays and normal probability plots. Also employed were UTM coordinates and data provided by the Argos Data Collection and Location Service.

It is worth mentioning again that this is work in progress. There are data from other animals; there is objective fitting to be done and model refinements to consider.

**Successive distances from La'au Point for the synthetic journey**

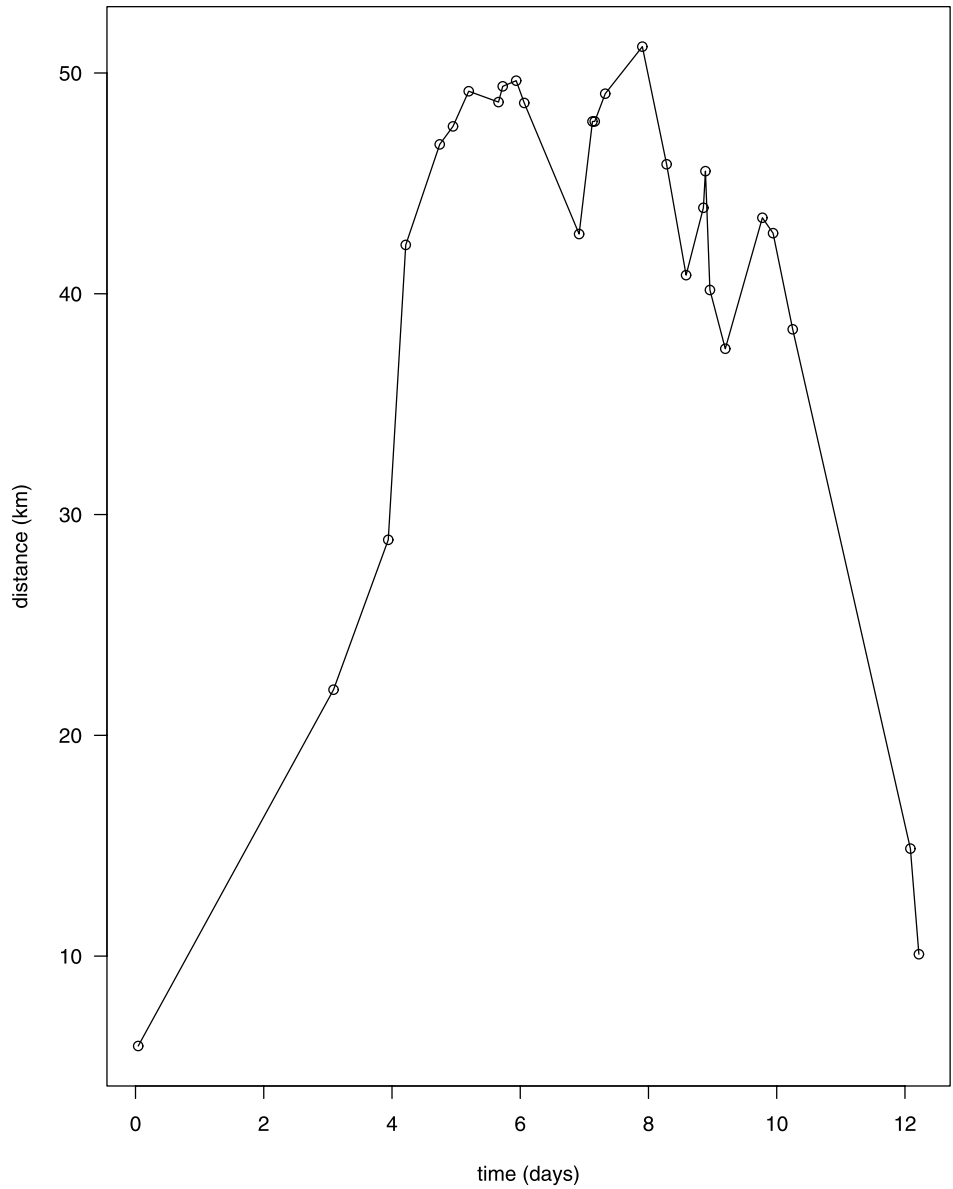

FIG 12. *Distances derived from the simulation of Figure 11. The times of the points plotted are those of journey 7 of Figure 7.*

There is great variability in foraging behavior between individuals. Perhaps development and application of this model will help us to be better able to determine which animals are doing different things and why. Nonetheless, the investigation and application of these models provides several insights into the foraging ecology of Hawaiian monk seals:

1. Particular marine habitats appear to be powerful attractors to foraging monk seals. In this case, these important foraging habitats are confined to relatively shallow offshore bathymetric features (i.e. less than 200 fathoms deep - Penguin Bank),
2. The time that the seal spent foraging there appeared to be constrained by another powerful attractor associated with periodic resting ashore (i.e., terrestrial haulout habitat).

### *À la prochaine*

"Freedman" and "Brill" have known each other since 1959. There was an initial suspicion, the one being from Montreal and the other from Toronto, but the relationship moved along. Best wishes to you, Freedman. May you find some Fest in this Schrift.

**Acknowledgments.** D. R. Brillinger thanks Ryan Lovett and Phil Spector of his Department for guidance with some knotty Linux and R problems.